\documentclass[preprint2]{aastex}

\newcommand{\myemail}{john@physics.uwa.edu.au}
\newcommand{\hiranoemail}{h-koichi@guitar.ocn.ne.jp}

\shorttitle{Redshift abundance periodicity in galaxy number counts}
\shortauthors{Hartnett}

\begin{document}

\title{Galaxy redshift abundance periodicity from Fourier analysis of number counts $N(z)$ using SDSS and 2dF GRS galaxy surveys}

\author{J.G. Hartnett}
\affil{School of Physics, the University of Western Australia\\35 Stirling Hwy, Crawley 6009 WA Australia; \myemail}
\author{K. Hirano}
\affil{Department of Physics, Tokyo University of Science\\1-3 Kagurazaka Shinjuku-ku, Tokyo 162-8601, Japan; \hiranoemail }

\begin{abstract}
A Fourier analysis on galaxy number counts from redshift data of both the Sloan Digital Sky Survey and the 2dF Galaxy Redshift Survey indicates that galaxies have preferred periodic redshift spacings of $\Delta z =$ 0.0102, 0.0246, and 0.0448 in the SDSS and strong agreement with the results from the 2dF GRS. The redshift spacings are confirmed by the mass density fluctuations, the power spectrum $P(z)$ and $N_{pairs}$ calculations. Application of the Hubble law results in galaxies preferentially located on co-moving concentric shells with periodic spacings. The combined results from both surveys indicate regular co-moving radial distance spacings of $31.7 \pm 1.8$ $h^{-1} Mpc$, $73.4 \pm 5.8$ $h^{-1} Mpc$ and $127 \pm 21$ $h^{-1} Mpc$. The results are consistent with oscillations in the expansion rate of the universe over past epochs. 
\end{abstract}



\section{\label{sec:Intro}Introduction}
When modeling the large scale structure of the cosmos the cosmological principle is assumed, therefore what we see must be biased by our viewpoint. The assumption is that the universe has expanded over time and any observer at any place at the same epoch would see essentially the same picture of the large scale distribution of galaxies in the universe. 

However evidence has emerged from both the 2dF Galaxy Redshift Survey \citep{2dFGRS} (see fig. 17 of Colless et al \citep{Colless}) and also the Sloan Digital Sky Survey \citep{SDSS} (see fig. 2 of Tegmark et al \citep{Tegmark2}) that seem to indicate that there is periodicity in the abundances of measured galaxy redshifts. This has emerged from the galaxy number counts ($N$) within a small redshift interval ($\delta z$) as a function of redshift ($z$).  

What has been found is the so-called \textit{picket-fence} structure of the $N$-$z$ relation, first noticed in one dimension by Broadhurst et al \citep{Broadhurst} from a pencil-beam survey of field galaxies. In this paper the galaxy surveys are analysed on 2D and 3D scales. The usual interpretation is that this is evidence of the Cosmic Web, including voids and long filaments of galaxies. An alternative interpretation is suggested in this paper; galaxies are preferentially found in redshift space at certain redshifts with higher number densities than at other redshifts. Of course, this does not rule out cosmic web structures in addition to this effect.

However the concept that this represents real space galaxy distances is incompatible with the cosmological principle, which assumes the uniformity of space at all epochs on sufficiently large enough scales.  And it has been demonstrated \citep{Yoshida}, from a large number of numerical simulations using the Einstein-de Sitter and $\Lambda$CDM models, that the probability of getting such a periodic spatial structure from clustering and Cosmic Web filaments is less than $10^{-3}$. This allows for an alternate explanation, namely that the redshift periodicity is evidence for past oscillations in the expansion rate of the universe.

The only information that has been sought from large scale galaxy surveys is the spatial power spectrum that describes the random yet uniform distribution of galaxies in the universe, where the only departure expected from the random distribution is some characteristic scale for the clustering of galaxies. Certainly very few expect to find evidence consistent with a periodic redshift distribution. 

However Hirano et al \citep{Hirano} have attempted to explain galaxy redshift abundance periodicity as a real effect resulting from the universe having a non-minimally coupled scalar field that manifests, in later cosmological times. The effect produced oscillations in the expansion rate as a function of time. They model not only the redshift space observations but also the effect their cosmology has on the CMB power spectrum and the type Ia supernova Hubble diagram.

In the usual analysis the spatial two-point or autocorrelation function is used to define the excess probability, compared to that expected for a random distribution, of finding a pair of galaxies at a given separation \citep{Baugh}. As a result the power spectrum $P(k)$ is derived from the two-point correlation function \citep{Peebles}.  The power spectrum is predicted by theories for the formation of large scale structure in the universe and compared with that measured, or more precisely calculated from the available data. 

The power spectrum for the SDSS has been calculated using a set of basis functions defined in \citep{Hamilton2}. Three power spectra (galaxy-galaxy, galaxy-velocity, and velocity-velocity) were determined. Local motions of galaxies only contribute a radial component of a galaxy's total redshift, hence only affect the radial component of the power spectrum. Because the angular power spectrum is unaffected the assumption is also made that galaxy-galaxy spectrum  is equal to the redshift-space power spectrum in the transverse direction. In this paper we analyze the redshift hence radial component of the power spectrum $P(z)$, calculated in redshift space not $k$-space, and find periodicity consistent with that observed in $N(z)$.  Also an $N_{pairs}$ analysis is carried out on both survey data sets with confirming results. 

The ``Bull's-eye'' effect has been studied \citep{Praton, Melott, Thomas} via $N$-body simulations to result from large-scale infall plus small-scale virial motion of galaxies. It is believed that these two effects can bias measurements of large scale real space galaxy distributions. One is the so-called ``fingers-of-God'' (FOG) effect, which results from random peculiar motions of galaxies at the same distance from the observer. The second acts on much larger scales and where overdensities of galaxies occur, like at the ``Great Wall'' for instance. See Fig. 1. 

Galaxies tend to have local motions toward the center of such structures. These motions are not random but coherent and add or subtract to the observer's line-of-sight redshift determination.  These effects preferentially distort the map in redshift space toward the observer due to the velocities of galaxies within clusters, i.e. non-cosmological redshift contributions. However the latter can only enhance in redshift space existing weak real space structures, it cannot create concentric structures centered on the Galaxy. And the FOG effect tends to smooth out finer detailed or higher Fourier frequency structures in  redshift space and is not very relevant to this discussion.

In the maps, fig. 18 of Colless et al \citep{Colless}) and fig. 4 of Tegmark et al \citep{Tegmark2}, there appears to the eye to be concentric structure in the arrangement of galaxies.  See the left hand side of the Fig. 1, the SDSS data with declinations between $0^{\circ}$ and $6^{\circ}$ mapped in Cartesian coordinates. Could it be that we inhabit a region of the Universe that is a bubble, part of a background of many such bubbles? But because redshift only gives us one dimensional information the conclusion could be drawn that if the galaxies in the Universe are evenly distributed throughout then redshift periodicity is indicative of oscillations in the expansion at past epochs. All observers would see the same shell structure in redshift space regardless of their location.

In this analysis we take a heuristic approach and look for periodicity in the redshift spacing between galaxies, in 2D in the case of slices and in 3D in case of the whole SDSS data set. In practice the discrete Fourier Transform ($\nu_s$) is calculated from the $N$-$z$ relations, which are histograms determined by binning (counting) the observed redshifts of the survey galaxies between $z-\delta z/2$ and $z+\delta z/2$, as a function of redshift $z$. Within the $i$th bin there are $\Delta N_i$ galaxies. Discrete Fourier amplitudes $\nu_s$ are generated for each integer $s>1$ from

\begin{equation} \label{FFT}
\nu_s =\frac{1}{\sqrt{n}}\sum_{i=1}^n u_s e^{2\pi j(i-1)(s-1)/n},
\end{equation}
where $u_s = \Delta N_i$, $j=\sqrt{-1}$, $n$ is the total number of $\Delta N_i$ bins.  Because $\nu_s$ are complex their absolute value is taken in the analysis. The Fourier frequencies are calculated from
\begin{equation} \label{Freq}
\mathcal{F} =\frac{n \delta z}{s} =\Delta z.
\end{equation}
where $\Delta z$ is the periodic redshift interval.

The results suggest that there are redshifts, on the scale size discussed here, which is not primarily the result of clustering or the FOG and infall effects resulting in distortion of the redshift space maps. Both the SDSS and 2dF GRS redshift data where the redshifts are known with 95\% confidence or greater are used. Fourier analysis can find periodic structure even where it may not be so apparent to the eye.
 
Data for 427,513 galaxies from the SDSS Fifth Data Release (DR5) were obtained where the data are primarily sampled from within about -10 to 70 degrees of the celestial equator. Also data for 229,193 galaxies were obtained from the 2dFGRS where the data are confined to within 2 degrees near the celestial equator and balanced between the Northern and Southern hemispheres. In the SDSS case the data are not so well balanced, yet there are nearly twice as many usable data. 

\section{Fourier analysis of $N$-$z$ relations}

\subsection{2D slices of SDSS data }
From the SDSS data three $6^{\circ}$ slices were taken in declination ($Dec$) angle for examination and comparison. They were taken from $0^{\circ} < Dec < 6^{\circ}$ , $40^{\circ} < Dec < 46^{\circ}$ and $52^{\circ} < Dec < 58^{\circ}$. A real space Cartesian plot (with aspect ratio of unity) of the data from the first slice near the equatorial plane is shown in Fig. 1. The redshifts have been converted to co-moving radial distance with Hubble constant $H_0 = 100 \, km s^{-1}Mpc^{-1}$ where the parameter $h = H_0/100$. The Great Wall is shown in the second and third quadrants as indicated. In those two quadrants it is evident to the eye that there is general concentric structure with a spacing of about 75 $h^{-1} \, Mpc$. 

Polar coordinate maps of these regions, projected onto the equatorial plane, are shown in Figs 2, 3 and 4, respectively. Most of the SDSS data come from two declination bands on the sky. The first slice is taken from the lower band near the celestial equator and the latter two from higher declinations. Polar maps can tend to distort a real space image but a comparison of Figs 1 and 2 indicates that these polar plots would be very close in appearance to real space maps if converted to co-moving coordinates. 

From simple inspection concentric structure is apparent in both Figs 2 and 4 but not as apparent in Fig. 3. To the eye this seems too coincidental and deserves further investigation. At least at some declinations the case can be made for redshift periodicity.
 
\begin{figure}
\includegraphics[width = 3 in]{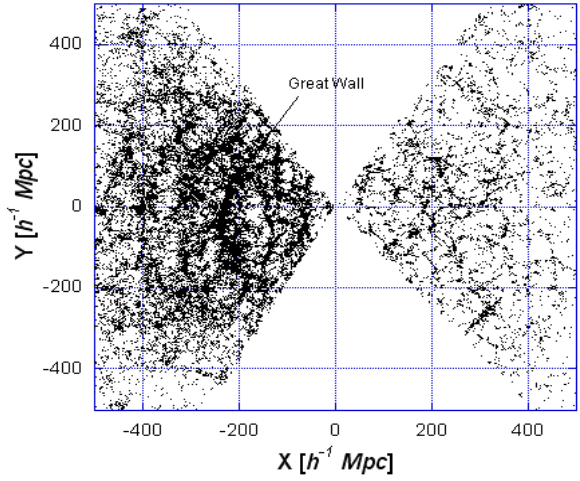}
\caption{\label{fig:fig1} Real space Cartesian coordinate plot in units of $h^{-1} \, Mpc$: co-moving (Hubble) distance derived from SDSS redshift data from $0^{\circ} < Dec < 6^{\circ}$. }
\end{figure}
\begin{figure}
\includegraphics[width = 3 in]{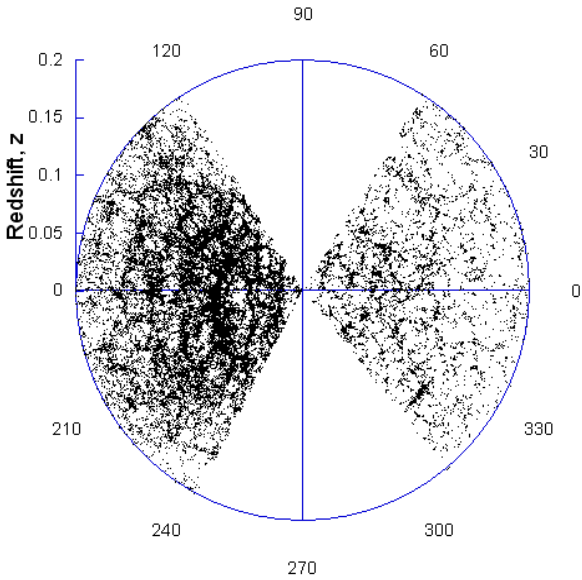}
\caption{\label{fig:fig2} Polar coordinate plot as a function of RA: Redshift space map of the same data as Fig. 1, taken  from  $0^{\circ} < Dec < 6^{\circ}$. There are 59,930 data for $z < 0.5$.}
\end{figure}
\begin{figure}
\includegraphics[width = 3 in]{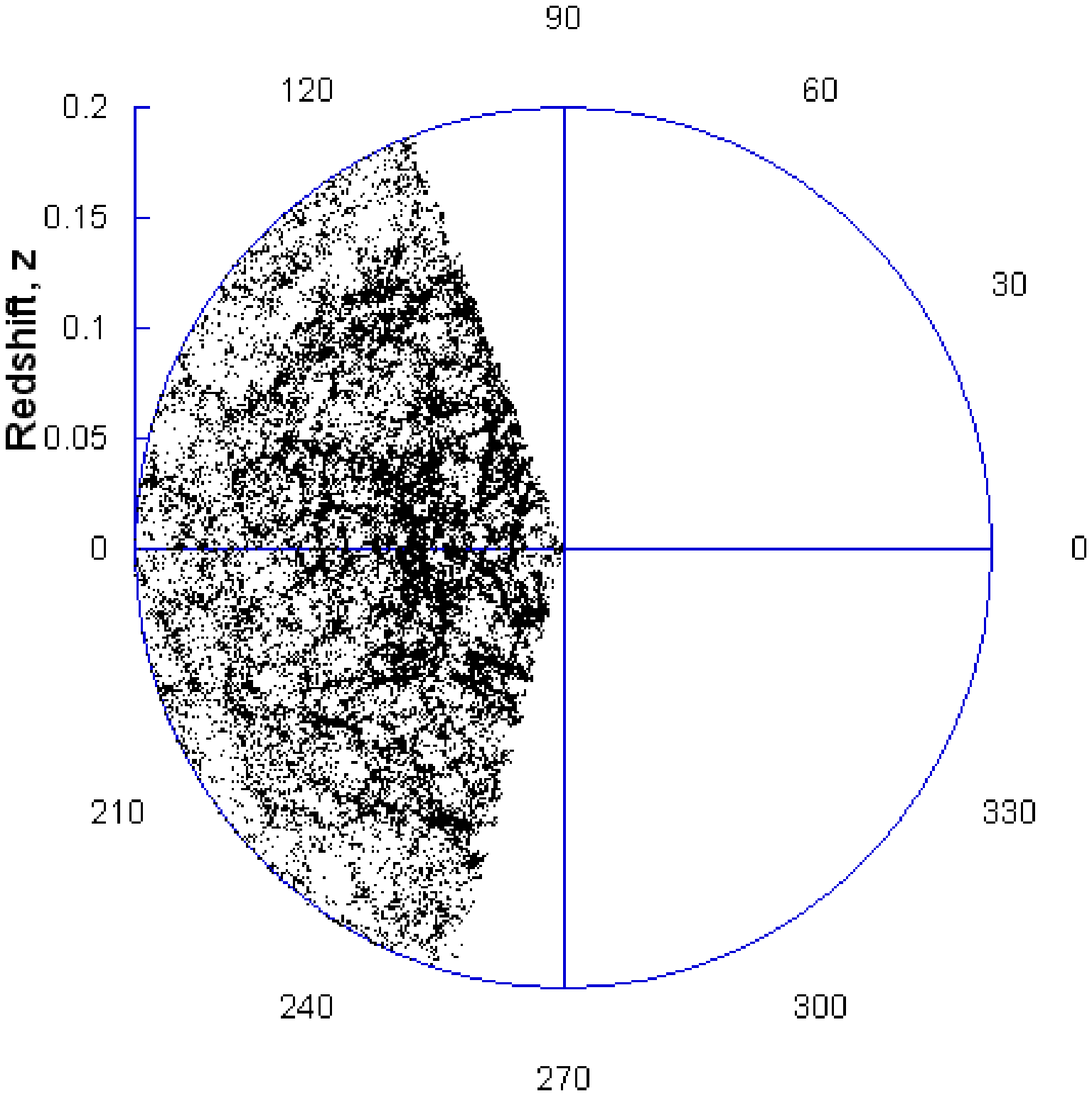}
\caption{\label{fig:fig3} Polar coordinate plot as a function of RA: Redshift space map of data from  $40^{\circ} < Dec < 46^{\circ}$. There are 44,335 data for $z < 0.5$. There are no observed data at these declinations in the two empty quadrants.}
\end{figure}
\begin{figure}
\includegraphics[width = 3 in]{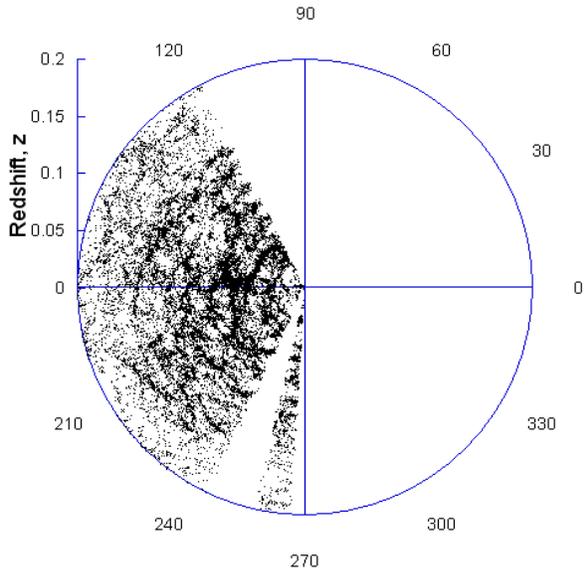}
\caption{\label{fig:fig4} Polar coordinate plot as a function of RA: Redshift space map of data from  $52^{\circ} < Dec < 58^{\circ}$. There are 34,802 data for $z < 0.5$. There are no observed data at these declinations in the two empty quadrants.}
\end{figure}
These data were binned with a bin size $\delta z = 10^{-3}$ resulting in a $N$-$z$ relation. The redshift resolution of most, but not all, of the SDSS data is only $10^{-3}$. This bin size essentially counts all galaxies of the same redshift regardless of their position on the sky. 

Using (\ref{FFT}) the discrete Fourier spectra were calculated for $N(z)$ of each data slice, and are shown in Fig. 5. The Fourier frequency axis in Fig. 5, and all subsequent figures, has been converted to a redshift interval ($\Delta z$). For the lowest angle slice (solid black curve) from the data of Fig. 2 and the highest angle slice (solid gray curve) from the data of Fig. 4 the peaks at numbers 1 and 2 coincide. For the data from the $40^{\circ} < Dec < 46^{\circ}$ slice of Fig. 3  the peaks are not strong nor do they coincide with the others except at peak number 1. The fact that the maps in Figs 1 through 3 are visually different  and that the resulting Fourier spectra don't all have coincidental peaks indicates that the peaks that do coincide are not the result of some selection effect resulting from the choice of filters or otherwise. Therefore it is evident that the peaks are the result of real redshift structure that varies in different directions. The underlying cause must be common to all, since all three slices have a Fourier peak at number 1, which represents a redshift space interval of $\Delta z \approx 10^{-2}$. 

\begin{figure}
\includegraphics[width = 3 in]{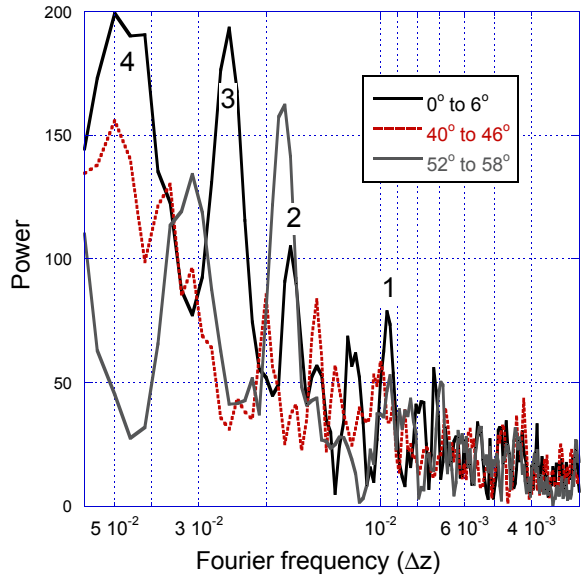}
\caption{\label{fig:fig5} Fourier transform of the $N$-$z$ data for the three $6^{\circ}$ declination slices on the sky. The black solid curve is for the $0^{\circ} < Dec < 6^{\circ}$ slice, the red broken curve is for the $40^{\circ} < Dec < 46^{\circ}$ slice and the gray solid curve is for the $52^{\circ} < Dec < 58^{\circ}$ slice.   The numbers are the positions of the peaks in the Fourier spectrum when all the SDSS data are taken together. See Fig. 7.}
\end{figure}

\subsection{3D: All SDSS data}
In order to see if large scale 3D concentric structure is evident the whole SDSS galaxy data set was binned and Fig. 6 shows the resulting $N$-$z$ relation for all $z \leq 0.35$. The discrete Fourier spectrum was calculated from $N(z)$ using (\ref{FFT})  and the result is shown in Fig. 7. It was found that $z > 0.15$ have very little effect on the result. 

\begin{figure}
\includegraphics[width = 3 in]{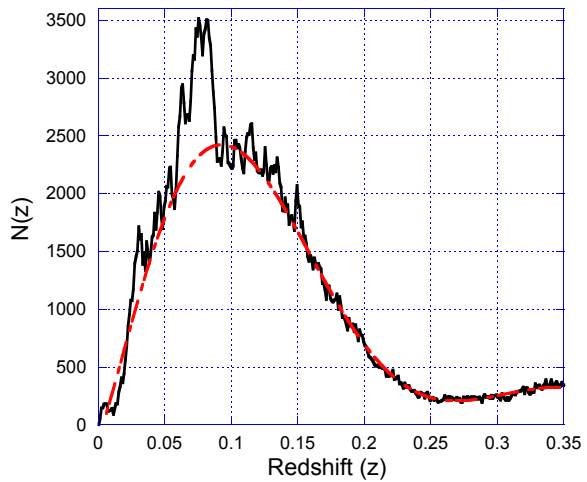}
\caption{\label{fig:fig6} The $N$-$z$ relation (solid black curve) from all the SDSS galaxy data $z \leq 0.35$, with redshift bins $\delta z = 10^{-3}$. The dot-dashed (red) curve is a polynomial fit for the survey function.}
\end{figure}
\begin{figure}
\includegraphics[width = 3 in]{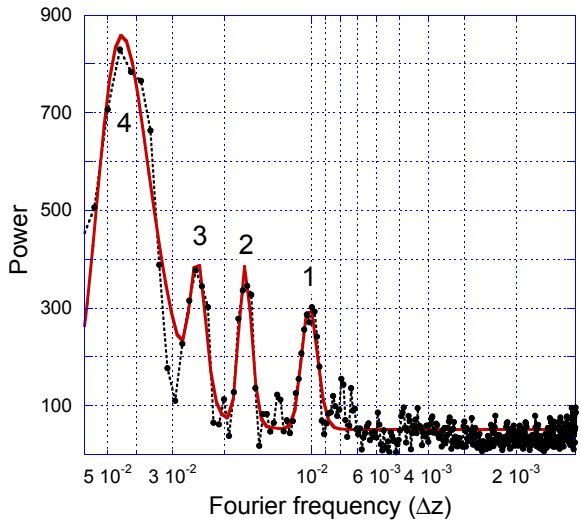}
\caption{\label{fig:fig7} The Fourier spectrum (filled circles, dotted line) of the SDSS galaxy $N$-$z$ data (curve in Fig. 6) where the DC Fourier component has been removed. The sum of 4 Gaussians has been fitted to the spectrum (solid black line).}
\end{figure}

Also the SDSS data were divided into two sets sampled from two opposite regions of the sky. The data are 374,425 with right ascensions $100^{\circ} < RA < 270^{\circ}$, and 53,087 with $290^{\circ} < RA < 50^{\circ}$. There is about 7 times the number of data in one set than the other. The larger set is mostly  from above the celestial equator.

$N(z)$ relations for these regions are shown in Fig. 8 on a double-Y axis since they are about 7 times different in scale. Nevertheless the general shape of the curves are the same indicating that $N(z)$ has the same $z$-dependence in each region. The location of peaks on these curves are similar except peaks occur with different relative heights and some at different redshifts. This indicates that though $N(z)$ has the same general dependence (shape of the galaxy survey function) the different regions are distinctly different in other respects.  Again this is evidence that the  $N$-$z$ relations are not artifacts of the filters or due to some random effects.

\begin{figure}
\includegraphics[width = 3 in]{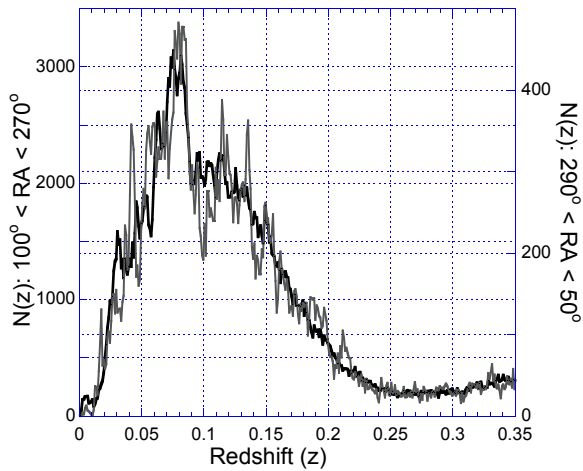}
\caption{\label{fig:fig8} $N$-$z$ data (with bins $\delta z = 10^{-3}$) from both regions of the sky. The black curve is from the larger region of the sky with $100^{\circ} < RA < 270^{\circ}$ and gray curve  is from the smaller region of the sky with $290^{\circ}  < RA < 50^{\circ}$.}
\end{figure}
\begin{figure}
\includegraphics[width = 3 in]{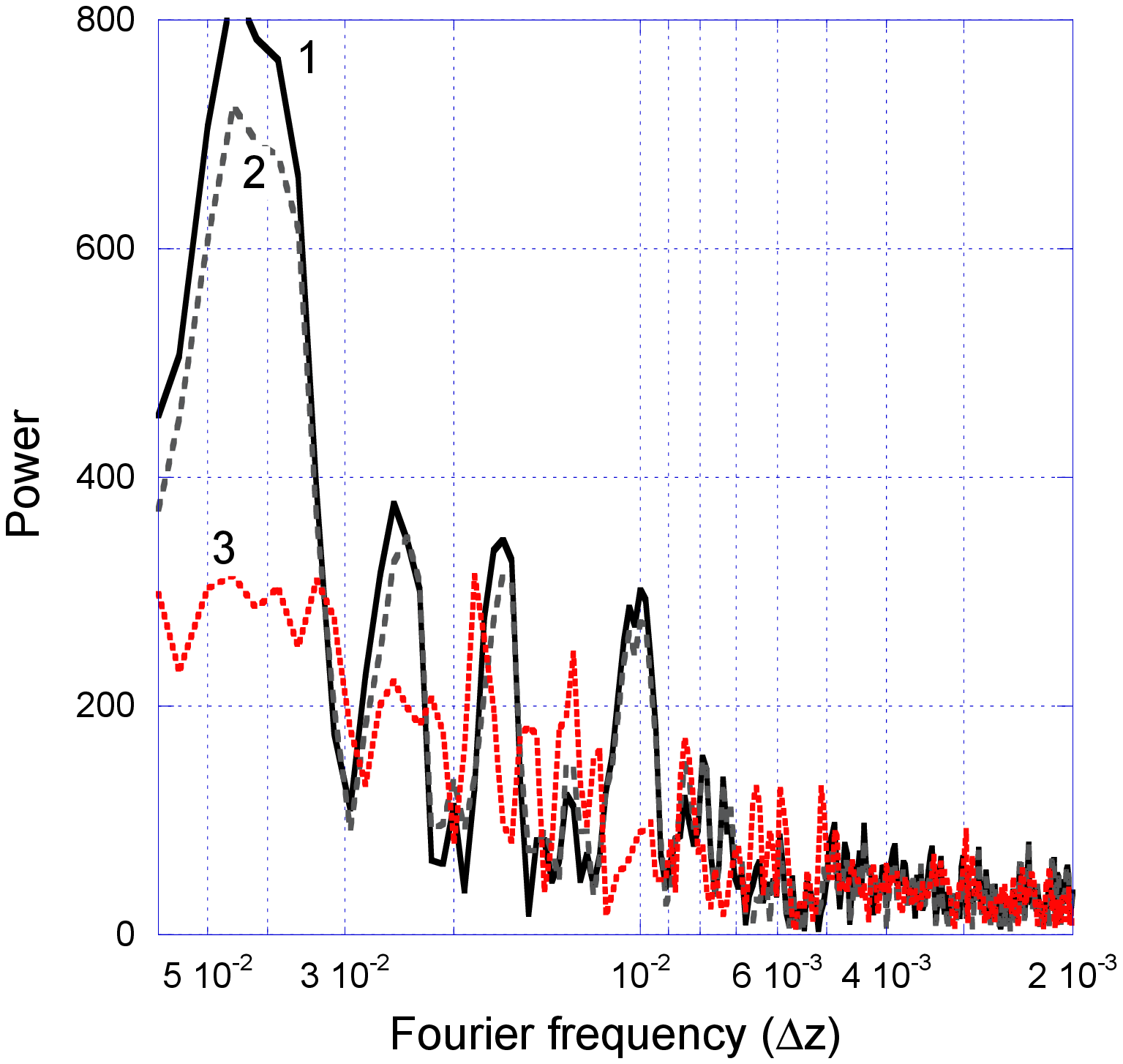}
\caption{\label{fig:fig9} Fourier transform of the $N$-$z$ relation  from Fig. 8. The solid (black) curve 1 is $N$-$z$  for the whole SDSS data set, the dashed (gray) curve 2 is for the larger region of the sky with $100^{\circ} < RA < 270^{\circ}$ and the dotted (red) curve 3 is for the smaller region of the sky with $290^{\circ} < RA < 50^{\circ}$, multiplied by a factor of 3.}
\end{figure}

This is seen also in their respective Fourier transforms as shown in Fig. 9. Curve 1 is the result from the whole SDSS data set (as shown in Fig. 7), curve 2 is the Fourier spectrum from the larger region of the sky surveyed with $100^{\circ} < RA < 270^{\circ}$ and curve 3 (enhanced by a factor of 3 for better comparison) is from the smaller region of the sky with $290^{\circ}  < RA < 50^{\circ}$. There are similar features but much weaker in the latter, yet also it is apparent that the peaks are shifted in Fourier frequency. This could indicate real differences in the redshift space structure in that part of the survey as compared to the other.

In order to evaluate the level of significance of the peaks from the whole SDSS data set the following was done. To the galaxy survey function, the smooth polynomial that fits the $N$-$z$ data of Fig. 6 but ignores the peaks (the dot-dashed curve in Fig. 6), white noise of the same magnitude as the variations of $N(z)$ at higher redshifts was added. Its Fourier spectra was then calculated and compared with the original in Fig. 10. This result indicates, as expected, that random white noise produces no peaks in the Fourier spectrum, but also that the 4 prominent peaks are anomalous in relation to what might be expected from a truly random distribution of galaxies. It is well known that galaxies cluster and also form super clusters, long filaments and voids, so some regularity could arise from this. From Fig. 10 it is apparent that the peaks are significant to at least 4$\sigma$, when compared to the simulated white noise spectrum.

\begin{figure}
\includegraphics[width = 3 in]{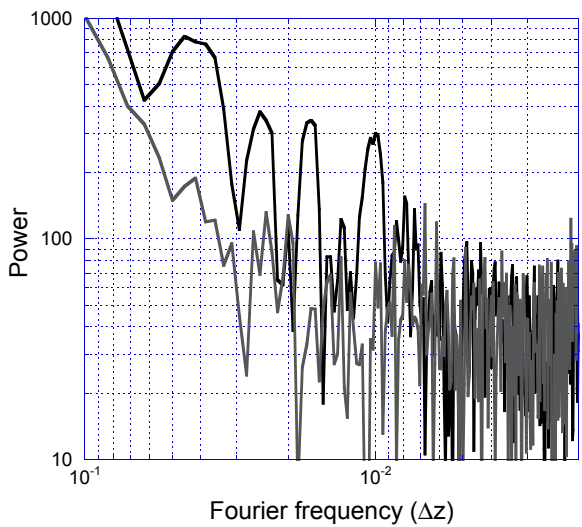}
\caption{\label{fig:fig10} The black curve is the discrete Fourier spectrum of the SDSS galaxy $N$-$z$ data (same as the solid curve in Fig. 7) plotted as a function Fourier frequency converted to $\Delta z$ interval. The gray curve is the Fourier spectrum of the white noise enhanced SDSS galaxy $N$-$z$ data drawn from the polynomial fit in Fig. 6.}
\end{figure}

\begin{table}[ph]
\begin{center}
Table I: Redshift ($\Delta z$) and distance intervals ($\Delta r$) for $H_0 = 100 \, km s^{-1}Mpc^{-1}$ from $N(z)$\\
\vspace{6pt}
\begin{tabular}{c|c|cccc} \hline
\hline
Survey										&Peak				&$\Delta z$ 	& error 	&$\Delta r$	& error\\
\hline
\hline
SDSS 											&1					&0.0102 			&0.0009			&30.6		&2.7\\
Fig. 7 										&2 					&0.0170     	&0.0012 		&51.0		&3.6\\
													&3					&0.0246 			&0.0023 		&73.8		&6.9\\
													&4					&0.0448				&0.0131 		&134		&39\\
\hline
2dFGRS 										&1					&0.0109 			&0.0008 		&32.7		&2.4\\
Fig. 12										&2 					&0.0145 			&0.0004 		&43.5		&1.2\\
													&3					&0.0244 			&0.0035 		&73.2		&10.5\\
													&4					&0.0415 			&0.0082 		&125		&25\\
\hline
\hline
\end{tabular}
\end{center}
\end{table}

The whole SDSS galaxy $N$-$z$ data Fourier spectrum is shown in Fig. 7 with the DC component removed. The sum of 4 Gaussian curves was fitted to this. The best fit peak $\Delta z$ values (representing preferred redshift intervals) and their standard deviations are shown in Table I. Using $H_0 = 100 \, km s^{-1} Mpc^{-1}$ the preferred redshift intervals (numbered peaks in Figs 7 and 12) were converted to co-moving distances and are listed in Table I.

\subsection{2D: SDSS and 2dF GRS compared}
In this section the SDSS data are compared with those from the 2dF GRS. Reliable redshifts were obtained for 229,193 galaxies. An $N$-$z$ relation calculated as a function of redshift $z$ was produced in Fig. 17 of Colless et al \citep{Colless}. This is reproduced in Fig. 11 from 2dF GRS data. The difference however is that the SDSS data cover a larger region of the sky than the narrow 2 degrees of the 2dF GRS data.

Nevertheless the Fourier spectrum of its $N(z)$ data shows very similar peaks to that of the SDSS data. See Fig. 12 where the peaks have been fitted with Gaussians, and Fig. 13 where the Fourier spectra from the SDSS and 2dF GRS data are compared. The values for the numbered peaks, representing the preferred redshift intervals ($\Delta z$), are listed in Table I, along with their standard errors. Fig. 13 shows the Fourier spectra from the whole SDSS data set (solid black curve), the whole 2dF GRS data set (dotted curve) and a subset of 49,044 SDSS data where the declination angle is within two degrees of the celestial equator (dashed curve). The map of the data for the latter closely resembles that of the 2dF GRS data as both are sampled close to the equatorial plane, and over the same right ascension.

From Fig. 12 it is clear that peak number 2 is not very significant in 2D samples. Its magnitude is similar to the noise at the higher Fourier frequencies. Notice though that the Fourier spectrum of the 2dF GRS data (dotted curve) is similar to the narrow equatorial slice from the SDSS data (dashed curve). Also the real space map of 2dF GRS data seen in Fig. 18 of Colless et al \citep{Colless} is very similar to Figs 1 and 2. Both show similar arcs and filament structures. 
\begin{figure}
\includegraphics[width = 3 in]{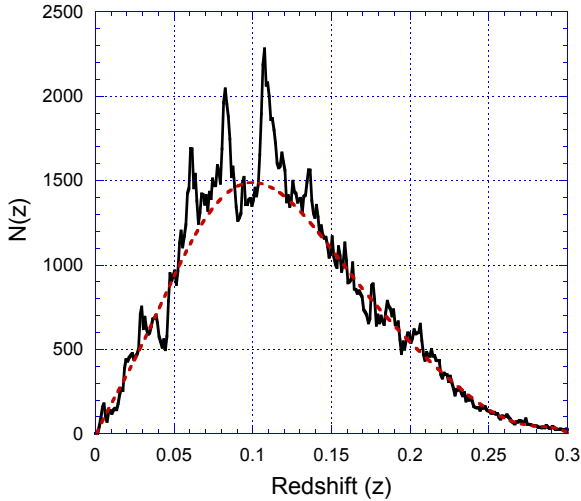}
\caption{\label{fig:fig11} The $N$-$z$ relation, with redshift bins $\delta z = 10^{-3}$, generated from the data for 229,193 galaxies in the 2dF GRS data release (see Fig. 17 of Colless et al \citep{Colless}). The dashed (red) curve is a polynomial fit for the survey function.  }
\end{figure}
\begin{figure}
\includegraphics[width = 3 in]{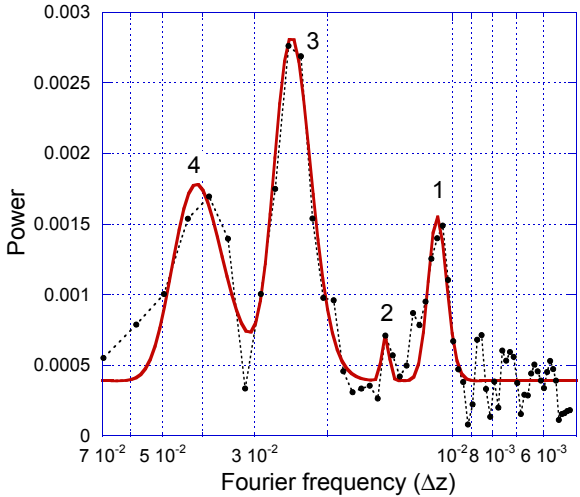}
\caption{\label{fig:fig12} Fourier transform of the $N$-$z$ data from the 2dF GRS (Fig. 11) with DC component removed.  The four lowest Fourier frequency peaks are shown. The peak data are shown in Table I. }
\end{figure}
\begin{figure}
\includegraphics[width = 3 in]{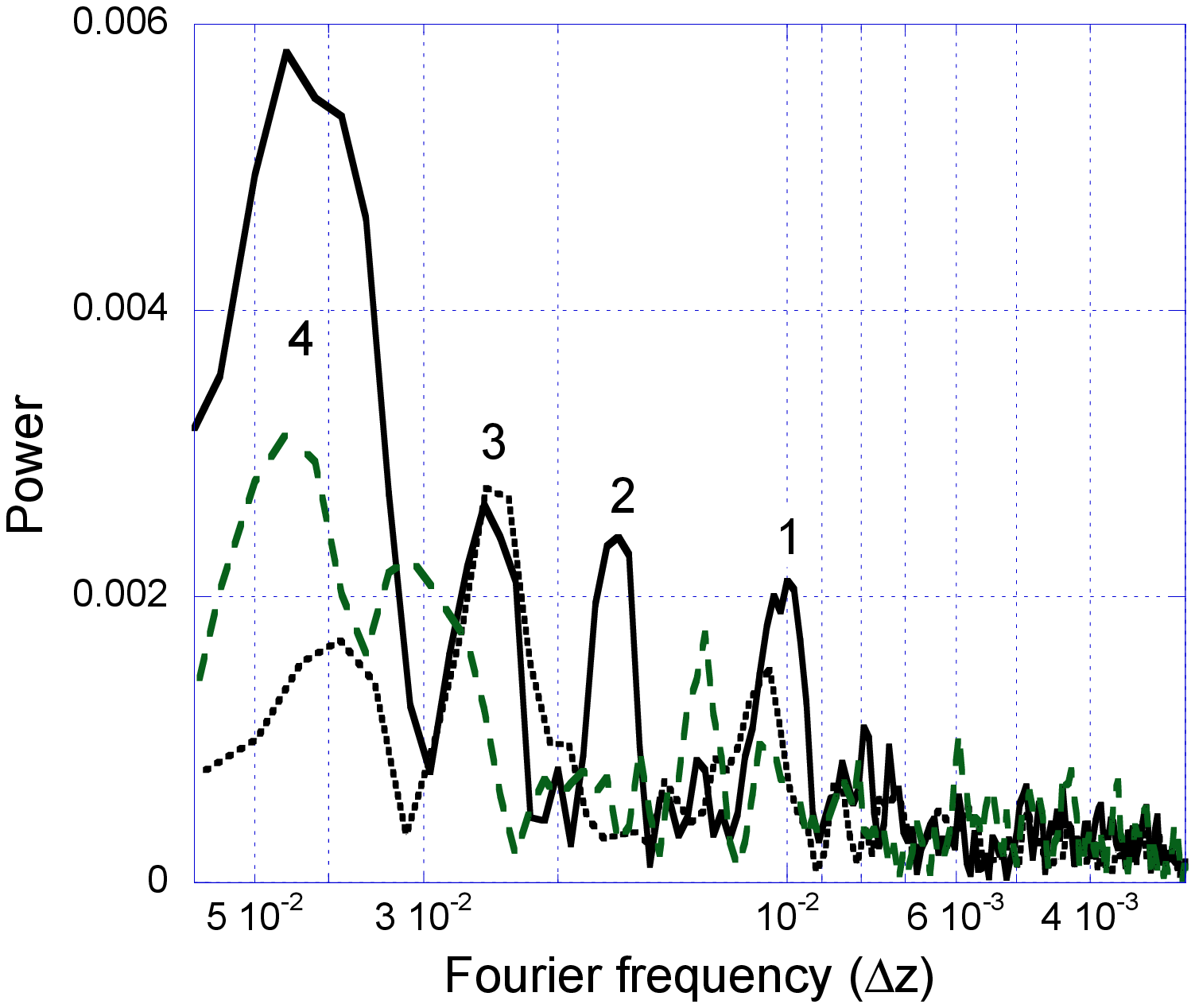}
\caption{\label{fig:fig13} Fourier transforms of the normalized $N$-$z$ data compared. The solid curve is from all the SDSS (Fig. 7), the dashed (green) curve from a $2^{\circ}$ slice through the equatorial plane from SDSS and the dotted curve is  from the 2dF GRS (Fig. 11). For the comparison the FTT spectrum from the SDSS has been normalized to the peak height of peak number 3 in the 2dF GRS result. }
\end{figure}

However peaks 1, 3 and 4 in Fig. 12 are all very significant. Also from a comparison of the data in Table I, it is clear that the values of the peaks 1, 3 and 4 are consistent with the respective values from the other data set, within their standard errors. 

Fig. 13 shows the Fourier spectra from the two surveys and except for peak number 2, they overlap. The height of peak number 3 from the SDSS was normalized to match peak number 3 of the 2dF GRS, so they could be compared.  It is observed that the Gaussian fits are a very good match.  Peak 3 is at a redshift interval $\Delta z = 0.0246$, which is most readily visible by eye in Fig. 11. Peaks at number 4 have a large height difference but similar line shapes.

Peak number 2 is strongly present in the Fourier spectrum from the whole SDSS data set but not significant in the 2dF GRS data. The Fourier spectrum of the thin equatorial slice from SDSS (dashed curve in Fig. 13) is similar to that of the 2dF GRS result. Peak number 2 seems to be very weak in both samples taken near the celestial equator. This indicates that it is not due to insufficient statistics. Peak number 1 then represents $\Delta z = 0.0102$,  which is also significant in the thin equatorial slice. 

\section{Mass density fluctuations}

Another way to analyze this data is to calculate the mass density fluctuations as a function of radial distance from the center of the spherical distribution. The \textit{rms} density fluctuations can be written in terms of the power spectrum $|\delta_k|^2$,
\begin{equation} \label{densityfluct}
\left(\frac{\delta \rho}{\rho}\right)^2=\left \langle \delta(\vec{x})\delta(\vec{x})\right \rangle,
\end{equation}
where $\left \langle \cdots \right \rangle$ indicates and average over all space. See \cite{Kolb} pp. 329--336. The contribution to $(\delta \rho/\rho)^2$ from a given logarithmic interval in $k$ is
\begin{equation} \label{variance}
\Delta^2(k) \equiv \frac{1}{V}\frac{k^3 |\delta_k|^2}{2\pi^2},
\end{equation}
which is the variance or the fluctuations in power per logarithmic $k$-interval. The \textit{rms} mass fluctuations $(\delta M/M)$ on a given mass scale are related to this. 

We use spherical coordinates and calculate the average mass density in a spherical volume of radius $r=n \delta r$ where $n$ is an integer and $\delta r = 3 \, h^{-1} Mpc$. This involves the reasonable assumption that the number density of galaxies is proportional to the mass density $\rho(\vec{x})$. Therefore for a given radius $r$
\begin{equation} \label{mass variance}
\left(\frac{\delta \rho}{\rho}\right)_r=\left(\frac{\delta M}{M}\right)_r.
\end{equation}
Below we calculate the averaged density fluctuations as a function of radius $r$ where $\delta \rho$ is the deviation from the average density $\rho$,
\begin{equation} \label{averagedens1}
\left(\frac{\delta \rho}{\rho}\right)_r=\frac{\Delta N}{\Delta V} /\frac{N(r)}{V(r)} - 1,
\end{equation}
where $\Delta N$ is the number of galaxies within the $n$th incremental volume $\Delta V$ determined by taking successive radius increments $\delta r$. $N(r)$ is the total number of galaxies within the volume to $r$ and $V(r)$ is the total volume to $r$. $(\delta \rho/\rho)^2$ from (\ref{averagedens1}) is the variance $\Delta^2(r)$ evaluated at $r$. From (\ref{averagedens1}) we get
\begin{equation} \label{averagedens2}
\left(\frac{\delta \rho}{\rho}\right)=\frac{\Delta N}{\Delta \Omega r^2 \delta r} \frac{\Delta \Omega \delta r \sum_{i=1}^n(i\delta r)^2}{N(r)} - 1,
\end{equation}
where the subscript has been dropped, $\Delta \Omega = sin\theta \Delta \theta \Delta \phi$ is the solid angle subtended by the $n$th volume element, which is a constant for data sampled between fixed $RA$ and $Dec$ angles. Therefore,
\begin{equation} \label{averagedens3}
\left(\frac{\delta \rho}{\rho}\right)=\frac{\Delta N(n)}{N(n)} \frac{\sum_{i=1}^n i^2}{n^2} -1,
\end{equation}
where $r=n \delta r$ has been used, and $N(n)$ is now a function of the $n$th increment.

Since $z=(H_0/c) r$ it is equally valid to use redshift intervals $\delta z$ where $z=n\delta z$. For the radial increment chosen here $\delta r = 3 \, h^{-1} Mpc$ the redshift increment $\delta z = 10^{-3}$. It follows from (\ref{averagedens3}) that 
\begin{equation} \label{averagedens4}
\left(\frac{\delta \rho}{\rho}\right)=\frac{\Delta N(n)}{\sum_{i=1}^n \Delta N(i)} \frac{(1+n)(1+2n)}{6 n} -1,
\end{equation}
where $\Delta N(z)$ is now the number of galaxies in the $n$th redshift interval, where $z=n\delta z$.
\begin{figure}
\includegraphics[width = 3 in]{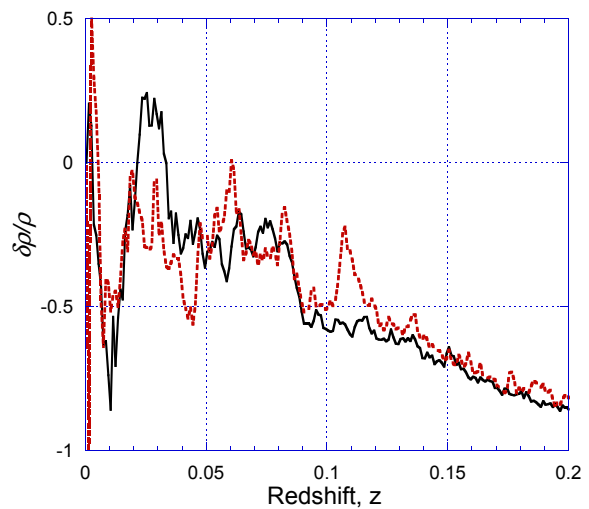}
\caption{\label{fig:fig14} $\delta \rho/\rho$ plotted as a function of redshift, $z$. The solid (black) curve is for all the SDSS data and the broken (red) curve for all the 2dF GRS data.}
\end{figure}
\begin{figure}
\includegraphics[width = 3 in]{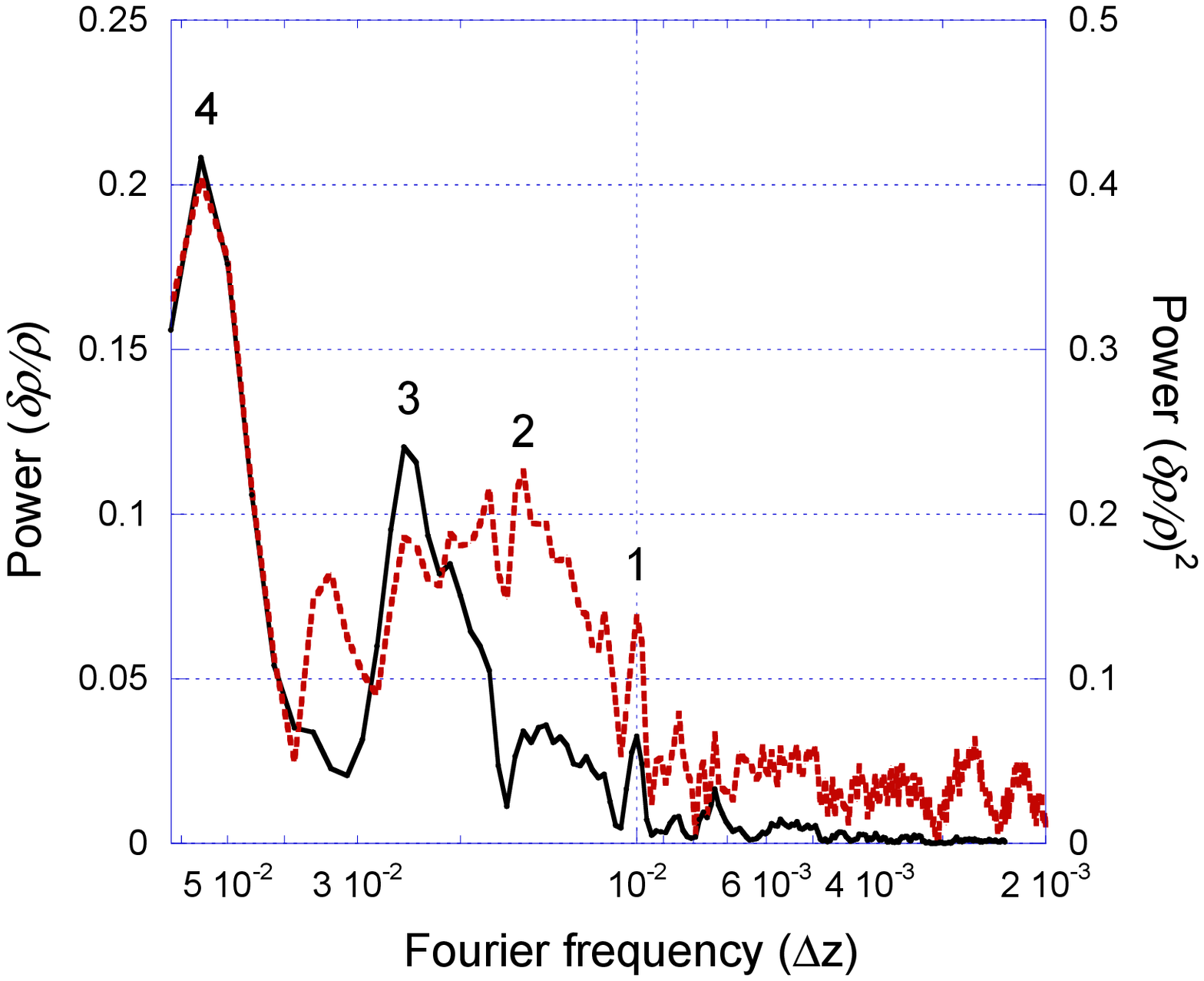}
\caption{\label{fig:fig15} The solid (black) curve is the squared Fourier transform of the SDSS $(\delta \rho/\rho)$ data from Fig. 14, plotted as a function of the redshift interval ($\Delta z$). The broken (red) curve is the Fourier transform of $(\delta \rho/\rho)^2$ derived from the same SDSS data. The most prominent peaks are labeled and correspond to the labels in Table I and II.}
\end{figure}
\begin{figure}
\includegraphics[width = 3 in]{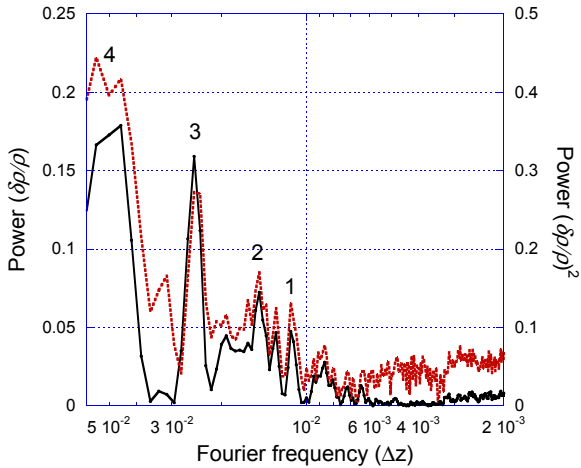}
\caption{\label{fig:fig16} The solid (black) curve is the squared Fourier transform of the 2dF GRS $(\delta \rho/\rho)$ data from Fig. 14, plotted as a function of the redshift interval ($\Delta z$). The broken (red) curve is the Fourier transform of $(\delta \rho/\rho)^2$ derived from the same 2dF GRS data. The most prominent peaks are labeled and correspond to the labels in Table I and II.}
\end{figure}

The resulting density fluctuations $(\delta \rho/\rho)$ have been plotted in Fig. 14 as a function of redshift for both the SDSS and 2dF GRS data. The square of these plots give you the variance $\Delta^2(z)$ from which the power spectrum $P(k)$ maybe derived with application of a windowing function and integrating over a logarithmic-$k$ interval. Here we calculate $P(z)$ in redshift units instead from
\begin{equation} \label{powerspec}
\left(\frac{\delta M}{M} \right)^2_{z_0} = \int^{\infty}_0 \Delta^2(z) W(z)^2 \frac{dz}{z},
\end{equation}
with constants set to unity and where $W(z)$ is the top-hat windowing function,
\begin{equation} \label{window}
W(z)=
\cases {1,\mbox{      }z \leq z_0 \cr
				0,\mbox{      }z > z_0. 		\cr} 
\end{equation}

The Fourier transforms of $(\delta \rho/\rho)$ and $(\delta \rho/\rho)^2$ were calculated using (\ref{FFT}) with $u_s = (\delta \rho/\rho)$ and $(\delta \rho/\rho)^2$, respectively. Fig. 15 shows the results for the SDSS data and Fig. 16 for the 2dF GRS data. The labeled peaks are listed  in Table II, where they have been calculated from Gaussian fits to the $(\delta \rho/\rho)$ curves. With the exception of Peak 4, the rest are all in good agreement with peaks determined from FFT analysis of the $N(z)$ counts in the previous section. 

\begin{table}[ph]
\begin{center}
Table II: Redshift ($\Delta z$) and distance intervals ($\Delta r$) for $H_0 = 100 \, km s^{-1}Mpc^{-1}$ from $(\delta \rho/\rho)$\\
\vspace{6pt}
\begin{tabular}{c|c|cccc} \hline
\hline
Survey										&Peak				&$\Delta z$ 	& error 	&$\Delta r$	& error\\
\hline
\hline
SDSS 											&1					&0.0101 			&0.0004			&30.3		&1.2\\
Fig. 15 									&2 					&0.0143     	&0.0036			&42.9		&10.8		\\
													&3					&0.0244 			&0.0044 		&73.2		&13.2\\
													&4					&0.0557				&0.0130 		&167  	&39\\
\hline
2dFGRS 										&1					&0.0112 			&0.0008 		&33.6		&2.4\\
Fig. 16										&2 					&0.0147 			&0.0012 		&44.1		&3.6\\
													&3					&0.0251 			&0.0020 		&75.6		&6.0\\
													&4					&0.0502 			&0.0110 		&151		&33\\
\hline
\hline
\end{tabular}
\end{center}
\end{table}

The power spectra $P(z)$ from all the SDSS and 2dF GRS redshift data are shown in Fig. 17. Both can be  closely approximated by $11.462 \, \Delta z$. Both power spectra deviate from the linear dependence and this can be seen in Fig. 18 where this linear dependence has been removed. The bottom (red) dotted curve is the difference between the two others. There is a large peak near $\Delta z_0 = 0.026$ and an indication of another near $\Delta z_0 =0.01$, indicated by the arrows. 

\begin{figure}
\includegraphics[width = 3 in]{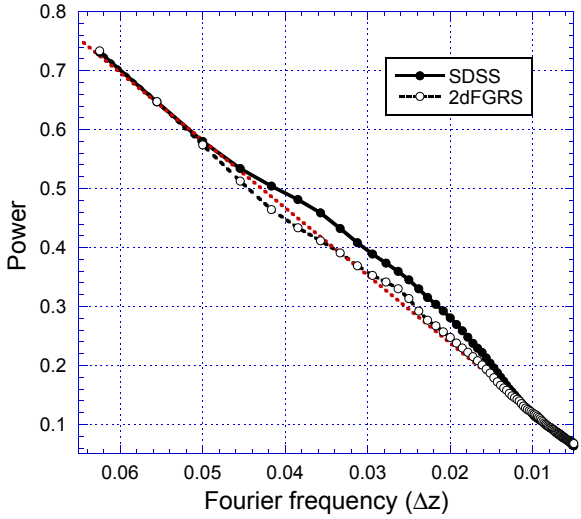}
\caption{\label{fig:fig17} Power spectra calculated from SDSS and 2dFGRS data. The dotted (red) curve is the fitted curve $9.012 \times 10^{-3} + 11.462 \, \Delta z$.}
\end{figure}
\begin{figure}
\includegraphics[width = 3 in]{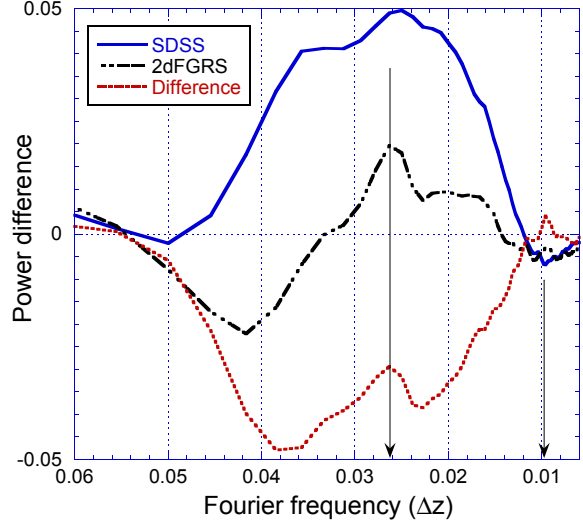}
\caption{\label{fig:fig18} Power spectra of Fig. 17 where the dotted (red) linear fit in Fig. 17 has been subtracted from both. The solid (blue) curve is for the SDSS data and the dot-dashed (black) curve for the 2dF GRS data. The dotted (red) curve is the difference between these two curves.}
\end{figure}

\section{Redshift separation between galaxies}

Another way to determine if galaxies are separated by a periodic spatial interval is to build histograms by binning the number of pairs of galaxies ($N_{pairs}$) that have the same  separation in redshift space ($\Delta z$). Then look for over abundance peaks in the histograms. Since redshifts are measured radially from the observer at the center of the distribution this method detects spatial separation with respect to that symmetry. 
\begin{figure}
\includegraphics[width = 3 in]{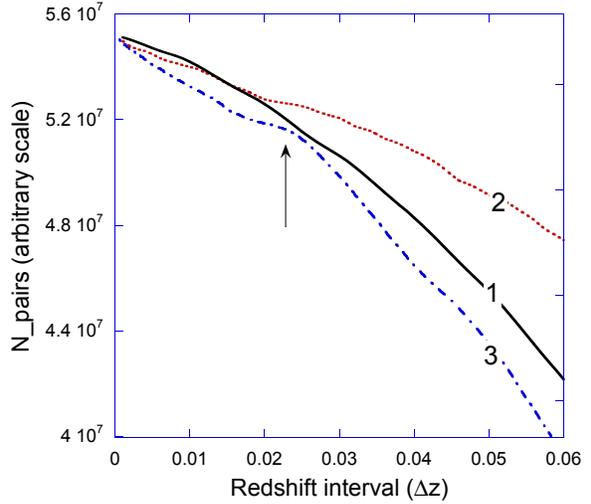}
\caption{\label{fig:fig19} The solid (black) curve (1) is the histogram for all the SDSS data, the dotted (red) curve (2) is the histogram for the SDSS slice and the dot-dashed (blue) curve  (3) is the histogram for all the 2dFGRS data. The arrow indicates a peak at $\Delta z \approx 0.025$.}
\end{figure}
\begin{figure}
\includegraphics[width = 3 in]{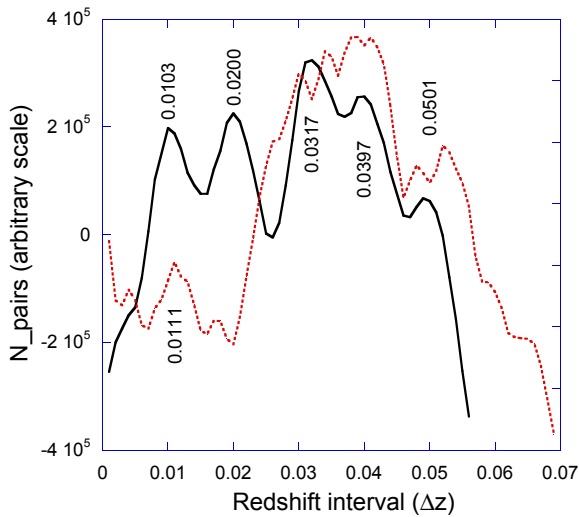}
\caption{\label{fig:fig20} The solid (black) curve is the histogram for all the SDSS data, the dotted (red) curve is the histogram for the SDSS slice after 2nd order polynomials have been subtracted to show fine detail of the peaks. The $\Delta z$ values for the peaks  are indicated. }
\end{figure}
\begin{figure}
\includegraphics[width = 3 in]{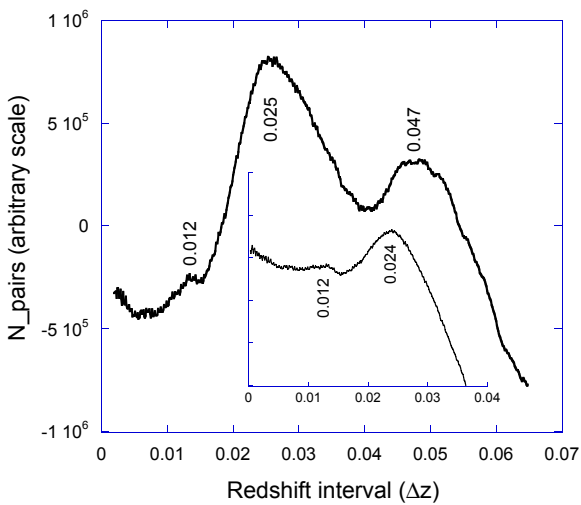}
\caption{\label{fig:fig21} The upper curve is the histogram for all the 2dF GRS data  after a 2nd order polynomial has been subtracted to show fine detail of the peaks. The inset lower curve is the same after an additional linear curve was subtracted to show the detail of the first peak. The $\Delta z$ values for the peaks are indicated. }
\end{figure}
\begin{figure}
\includegraphics[width = 3 in]{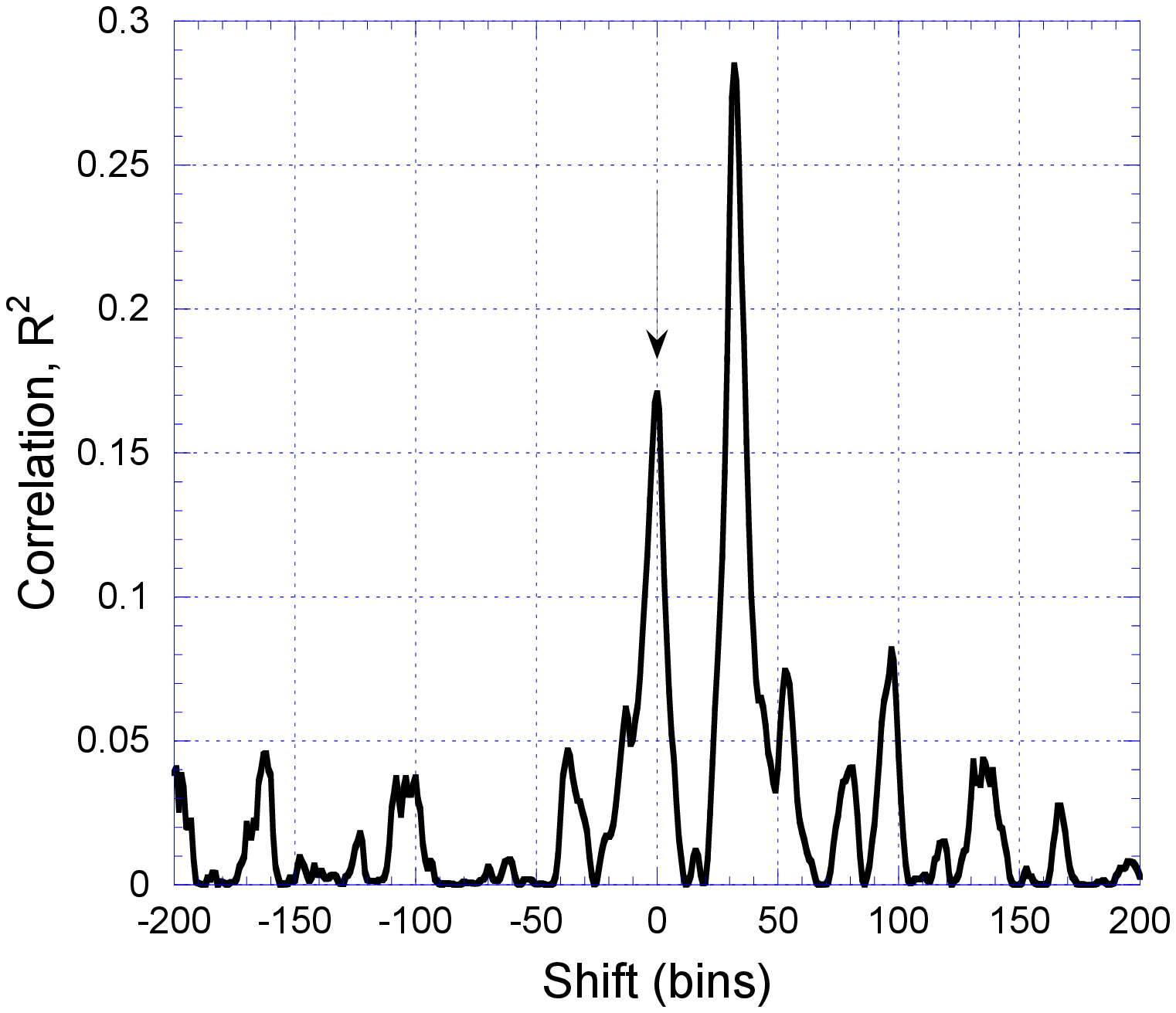}
\caption{\label{fig:fig22} The correlation function $R^2$ between the abundance distributions of SDSS and 2dF GRS. The zero shifted correlation coefficient is shown by the arrow.}
\end{figure}

Three histograms were built and are shown in Fig. 19:
\begin{enumerate}
	\item all the 427,513 SDSS data from DR5 release,
	\item the slice of 49,044 SDSS data taken from within $\pm 2^{\circ}$ of the celestial equator, and,
	\item all the 229,193  2dFGRS data taken within $2^{\circ}$ of the celestial equator with reliable redshifts.
\end{enumerate}

In Fig. 19 one can see slight ripples in these curves which represent above average correlations.  Clearly this method is not very sensitive to distinguish preferred separations against the background of many pairs that have no such correlation. However one peak is clearly visible as indicated by the arrow; this is at $\Delta z \approx 0.025$.

By subtracting off a second order (quadratic) polynomial from these curves one can separate off the background, enhance the weak features and see the peaks more clearly, as shown in Figs 20 and 21. Nevertheless they are still quite weak. In Fig. 20 the solid (black) curve for $N_{pairs}$ from all  the SDSS data shows peaks at $\Delta z= 0.0103$ in good agreement with the previous analyses, at $\Delta z= 0.0200$ and $0.0501$ in reasonably agreement,  but the others are not in agreement. The dotted (red) curve from the SDSS equatorial slice is quite noisy without any real clear peaks except the one near $\Delta z = 0.011$ as indicated. It is possibly the only reliable peak considering the subtraction process.

Fig. 21 shows the $N_{pairs}$ results from all the 2dF GRS data after a quadratic curve has also been removed to enhance the fine detail of the peaks. That is shown in the upper curve. The lower curve results after a further linear curve was subtracted to enhance the first peak. The results from the 2dF GRS $N_{pairs}$ analysis gives better results than that for the SDSS data. In the former case we see peaks around $\Delta z = 0.012, 0.025$ and $0.047$ in good agreement with the previous analyses.

\section{Correlation between two surveys}

The correlation between the $N(z)$ counts from the two surveys was evaluated after subtracting off the background galaxy survey functions, indicated by polynomial fits in Figs 6 and 11. This leaves only the abundance  peaks as a function of redshift which were used in the following correlation function.

\begin{equation} \label{correlation}
R^2=\frac{\left(n\sum X Y-(\sum X)(\sum Y) \right)^2}{[n\sum X^2-(\sum X)^2][n\sum Y^2-(\sum Y)^2]},
\end{equation} 
where $X$ and $Y$ represent $N(z)$ from the two surveys and $n$ is the number of $\Delta N_i$ bins in the $N(z)$ distribution--the same number and bin width for each survey.

The function $R^2$ is calculated while incrementally shifting the bin number $i$, assigned to $\Delta N_i$ in either the $X$ or $Y$ distributions, by unity up to half the total number of bins. For strong correlation the unshifted bins (0) value for $R^2$ will be maximum (where $R^2 \leq 1$). The result is shown in Fig. 22, where the zero shifted value is indicated by the arrow. The unshifted value of $R^2$ is maximum except for the the single large peak to its right, which can be explained.

If both distributions are periodic with the same period we'd expect also periodic structure in the correlation function $R^2$, which is what we in fact see. The Fourier transform of $R^2$ has a single peak where $\Delta z \approx 0.027$. The strong peak on the right of the central 0-shifted value results when the other of the two central abundance peaks in Fig. 11 aligns with strong peak of Fig. 6.

\section{Discussion}
This analysis finds redshift spacings of $\Delta z =$ 0.0102, 0.0170, 0.0246, and 0.0448 in the SDSS and very similar results from 2dF GRS. The most significant peaks in the Fourier spectra, calculated from the galaxy $N$-$z$ data, represent the first, third and forth spacings. By combining results from both surveys we get the most probable values for regular co-moving distance spacings of $31.7 \pm 1.8$ $h^{-1} Mpc$, $73.4 \pm 5.8$ $h^{-1} Mpc$ and $127 \pm 21$ $h^{-1} Mpc$. Because the 2dF GRS sampled close to the celestial equator and the second peak is not prominent in the SDSS data when sampled near the celestial equator (see dashed curve in Fig. 13), the second peak must be related to some redshift symmetry at higher declinations. It is strongly seen in the SDSS $52^{\circ} < Dec < 58^{\circ}$ slice. See peak 2 in the solid gray curve in Fig. 5. Peak number 4 ($127 \pm 21 \, h^{-1} Mpc$) is in agreement with the results of  Broadhurst et al \citep{Broadhurst1988, Broadhurst}. 

The mass density as a function of redshift has been calculated from the available data from both surveys and the Fourier transforms calculated. The results are in good agreement with the Fourier transform of the $N$-$z$ data. Compare Tables I and II. The power spectrum is not very sensitive to these effects but nevertheless peaks are detected at the same redshift periods. Also pair-counts were done on both surveys and the $N_{pairs}$ histograms showed  (albeit weakly) some peaks. The best results were gained for the 2dF GRS data. However in the SDSS data that method is not sufficiently sensitive to detect the effect sought.

\section{Conclusion}
A lot of work has been done in the last 20 years trying to show that the spatial structure determined from galaxy survey data is in agreement with what is predicted by the standard model. The standard model is assumed to be correct and when evidence questioning that conclusion is found (Broadhurst et al. \cite{Broadhurst} and evidence for concentric shells) a special effort was immediately made to show how it could still be explained in the standard model \cite{Kaiser,Praton}. But Kaiser and Peacock also concluded that their results ``did not prove that unconventional pictures for large-scale structures were ruled out.''  The appearance of concentric rings, like the \textit{picket fence}, can be explained if appropriate assumptions are made (e.g. the Bull's Eye effect), but this too does not mean that that explanation is necessarily correct. Furthermore, surely one cannot attribute peaks in the power spectrum to a windowing function if the structures producing these peaks are already visible in the raw data, before the Fourier transform is taken. 

This paper uses a simple Fourier analysis of number counts of galaxy redshifts in the SDSS and 2df GRS galaxy samples to search for periodic redshift spacings. Several spacings are found and are also confirmed using mass density fluctuations and the redshift separations between galaxies, though the latter are less sensitive methods. This analysis is similar in many ways to recent analyses carried out by other investigators, such as Tegmark et al. \cite{Tegmark2} where the power spectra obtained from the SDSS and 2dF GRS samples are also found to have similar power peaks.

Acoustic oscillations in the plasma during the pre-recombination epoch are expected to have been imprinted on the distribution of matter in the Universe and to have survived until the present. In standard model the size, distribution, and clustering of galaxies were set in the very early Universe. In addition to the windowing function, non-linear evolution of structure and scale dependent bias, also play a role in our correct understanding of this. In this paper we have introduced another possible uncertainty into these studies that has not previously been considered. We suggest that the power peaks we find are also consistent with oscillations in the expansion rate at recent epochs. As unlikely as this may seem, it is not possible to rule it out completely since there is visible evidence in the raw data for an apparent concentric shell structure centered on the observer. It would be easier to rule it out if one could argue that such oscillations in the Hubble flow are impossible. However, astronomers are now arguing for a speed-up in the Hubble flow at the present epoch. Also a model was recently published (Hirano et al. \cite{Hirano}) describing the origin of such oscillations.

One of the arguments against the power peaks being related to oscillations in the Hubble flow is the fact that the $P(k)$ distribution is a reasonable fit to what is predicted for the standard model. However, this does not rule oscillations out also. If there were no density fluctuations (no clustering  etc.), and the Universe was made up of a completely uniform distribution of galaxies, oscillations in the raw Hubble flow data would likely result in the appearance of concentric shells of apparent low and high density surrounding the observer because of the light travel time. To detect this there may be no need to use a windowing function. However, even if oscillations are present in the low-redshift Universe, any accurate model would also have to account for the density fluctuations (due to clusters, super clusters, etc.) that are also clearly visible.

And some have argued that the peaks found in this paper are just the scale sizes of clustering \citep{Kaiser, Mo, Gonzalez}; this was considered improbable by \citep{Yoshida} but a similar analysis needs to be performed on a larger data set in order to test this. Nevertheless the methods applied to the data here are consistent with the Universe undergoing oscillations in its expansion rate over past epochs. This means in redshift space there are preferred redshifts where galaxies are more dense due to a slower expansion rate at those epochs.  

\section{Acknowledgments}
We would like to thank F. Oliveira and T. Potter for valuable discussions and assisting JGH in obtaining SDSS/2dF GRS data, as well as P. Abbott for help with Mathematica software. The concluding remarks were largely written from comments made by one anonymous reviewer, for which we are very thankful. This work has been supported by the Australian Research Council.

Funding for the SDSS and SDSS-II has been provided by the Alfred P. Sloan Foundation, the Participating Institutions, the National Science Foundation, the U.S. Department of Energy, the National Aeronautics and Space Administration, the Japanese Monbukagakusho, the Max Planck Society, and the Higher Education Funding Council for England. 

The SDSS is managed by the Astrophysical Research Consortium for the Participating Institutions. The Participating Institutions are the American Museum of Natural History, Astrophysical Institute Potsdam, University of Basel, University of Cambridge, Case Western Reserve University, University of Chicago, Drexel University, Fermilab, the Institute for Advanced Study, the Japan Participation Group, Johns Hopkins University, the Joint Institute for Nuclear Astrophysics, the Kavli Institute for Particle Astrophysics and Cosmology, the Korean Scientist Group, the Chinese Academy of Sciences (LAMOST), Los Alamos National Laboratory, the Max-Planck-Institute for Astronomy (MPIA), the Max-Planck-Institute for Astrophysics (MPA), New Mexico State University, Ohio State University, University of Pittsburgh, University of Portsmouth, Princeton University, the United States Naval Observatory, and the University of Washington.

\end{document}